\documentclass[pra,aps,twocolumn,floatfix,superscriptaddress,showpacs]{revtex4}
\usepackage{bm,graphicx,amsmath}
\usepackage{times,txfonts}

\newcommand{\rp}[1]{(\ref{#1})}

\newcommand{\abs}[1]{\left|{#1}\right|}

\newcommand{\av}[1]{\left\langle #1 \right\rangle}

\newcommand{\br}[1]{\langle #1|}

\newcommand{\ke}[1]{|#1\rangle}

\newcommand{\bk}[2]{\langle #1|#2\rangle}
\newcommand{\kb}[2]{\ke{#1}\br{#2}}

\newcommand{\da}{^\dagger}

\newcommand{\pt}[1]{\left( #1 \right)}
\newcommand{\pq}[1]{\left[ #1 \right]}
\newcommand{\pg}[1]{\left\{ #1 \right\}}

\newcommand{\lpq}[1]{\left[ #1 \right.}

\newcommand{\rpq}[1]{\left. #1 \right]}

\newcommand{\ee}{{\rm e}}
\newcommand{\ii}{{\rm i}}

\newcommand{\nn}{{\nonumber}}

\newcommand{\CC}{{\cal C}}

\begin{document}

 \title{Simulating long-distance entanglement in quantum spin chains by superconducting flux qubits}

\author{Stefano Zippilli}
\affiliation{\mbox{Dipartimento di Ingegneria Industriale, Universit\`a degli Studi di Salerno, Via Giovanni Paolo II 132, I-84084 Fisciano (SA), Italy}}
\affiliation{School of Science and Technology, Physics Division, University of Camerino, via Madonna delle Carceri, 9, I-62032 Camerino (MC), Italy, and INFN, Sezione di Perugia, Italy}

\author{Miroslav Grajcar}
\affiliation{Department of Experimental Physics, Comenius University, SK-84248 Bratislava, Slovakia}
\affiliation{Institute of Physics, Slovak Academy of Sciences, D\'{u}bravsk\'{a} cesta, Bratislava, Slovakia}

\author{Evgeni Il'ichev}
\affiliation{Leibniz Institute of Photonic Technology, P.O. Box 100239, D-07702 Jena, Germany}
\affiliation{Novosibirsk State Technical University, 20 Karl Marx Avenue, 630092 Novosibirsk, Russia}

\author{Fabrizio Illuminati}
\thanks{Corresponding author: fabrizio.illuminati@gmail.com}
\affiliation{\mbox{Dipartimento di Ingegneria Industriale, Universit\`a degli Studi di Salerno, Via Giovanni Paolo II 132, I-84084 Fisciano (SA), Italy}}
\affiliation{INFN, Sezione di Napoli, Gruppo collegato di Salerno, I-84084 Fisciano (SA), Italy}

\date{January 22, 2015}

\begin{abstract}
We investigate the performance of superconducting flux qubits for the adiabatic quantum simulation of long distance entanglement (LDE), namely a finite ground-state entanglement between the end spins of a quantum spin chain with open boundary conditions. As such, LDE can be considered an elementary precursor of edge modes and topological order. We discuss two possible implementations which simulate open chains with uniform bulk and weak end bonds, either with Ising or with $XX$ nearest-neighbor interactions. In both cases we discuss a suitable protocol for the adiabatic preparation of the ground state in the physical regimes featuring LDE. In the first case the adiabatic manipulation and the Ising interactions are realized using dc-currents, while in the second
case microwaves fields are used to control the smoothness of the transformation and to realize the effective $XX$ interactions. We demonstrate the adiabatic preparation of the end-to-end entanglement in chains of four qubits with realistic parameters and on a relatively fast time scale.
\end{abstract}



\maketitle

\date{October 17, 2014}

\section{Introduction}

The inextricable complexity of many body quantum systems can be efficiently analyzed with the aid of quantum simulators~\cite{Feynman,Lloyd}, namely quantum devices consisting of many interacting systems that can be used to engineer and reproduce, in a controlled way, the dynamics of complex quantum models.
Superconducting devices based on Josephson Junctions are extremely versatile systems that hold promise for the efficient implementation of qubits for quantum technology applications~\cite{Makhlin,Clarke,Devoret}; in particular, they have been proposed as one of the most promising platform for the implementation of quantum simulators~\cite{Kaminski04,Grajcar05, You,Johnson,Houck,Izmalkov06}.

\begin{figure}[!th]
\begin{center}
\includegraphics[width=8cm]{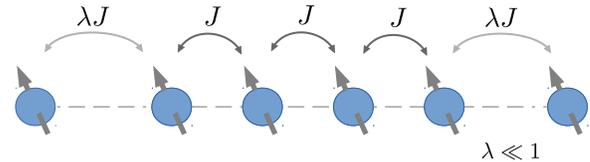}
\end{center}
\caption{An quantum spin chain with open boundary conditions and nearest-neighbor interactions featuring a uniform bulk with strong inter-spin coupling $J$ and two edge spins attached to the rest of the chain by weak end bonds $\lambda J$, with $\lambda \ll 1$.}
\label{figspinchain}
\end{figure}

Here we show how superconducting devices can be manipulated to simulate the phenomenon of long distance entanglement (LDE), namely the nonvanishing entanglement that is established between the end spins in the ground state of a quantum spin chain with open boundary conditions. The end points of the chain are in general non
directly interacting and in principle can be separated by arbitrary large distances. Nevertheless, they can become strongly entangled if, as shown in Fig.~\ref{figspinchain}, one implements specific patterns of interactions such as a strongly interacting uniform bulk coupled to the boundary spins by weak end bonds or a regular pattern of alternating weak and strong bonds~\cite{CamposVenuti1,CamposVenuti2,GiampaoloLong1,GiampaoloLong2}. LDE can be further enhanced by considering indefinitely repeated modular chains, giving rise to modular entanglement (ME)~\cite{Gualdi}, and is generalized to the so-called surface entanglement (SE) between distant and non-interacting spins belonging to the surface of  open two-dimensional networks with weak boundary-to-bulk coupling patterns~\cite{ZippilliSurf}.

LDE and its generalizations are potentially important concepts and tools because efficient schemes, such as quantum repeaters, for the distribution of entanglement between remote and non-directly interacting resources are essential to quantum information and communication applications~\cite{Cirac,Bose2003,Zippilli08,Zippilli2013,Roch}. In this context, realizing LDE in spin chains would be an efficient mean to operate and control distant qubits inside a quantum processor. Moreover, LDE, ME, and SE can be seen as elementary precursors of the role that boundary conditions and
edge modes can play in the physics of quantum many-body systems, anticipating some of the characteristic traits of topologically ordered phases and other exotic forms of nonlocal order; in particular, ground-state LDE in spin chains of the $XX$ and Heisenberg type with specific non-uniform coupling patterns is loosely reminiscent of more complex forms of nonlocal order such as symmetry-protected topological order~\cite{Wen,Pollmann}, which is realized in more elaborate one-dimensional models with uniform couplings and open boundary conditions, such as the spin-1 Heisenberg chain~\cite{Haldane,AKLT}, the cluster-Ising and cluster-$XY$ models with three-body local interaction terms~\cite{Smacchia,Hamma,Cui}, and the Kitaev fermionic chain with edge Majorana modes~\cite{Kitaev}.

The origin of LDE in a quantum spin chain with open boundary conditions can be understood as the effect of a strongly correlated bulk that mediates an effective entangling interaction between the two weakly coupled spins at the two ends of the chain. As already mentioned, this concept can be extended to higher-dimensional open spin networks\cite{ZippilliSurf}, for which the external spins on the surface of the network can be endowed with a rich variety of entanglement structures. This phenomenon has been identified in a large class of spin models ranging from the $XX$ to the $XXZ$ and the fully isotropic Heisenberg Hamiltonian~\cite{ GiampaoloLong1,GiampaoloLong2}. While in general LDE is not observed in open chains with Ising-type interactions without external field, the inclusion of a moderate transverse field can give rise to LDE, as discussed in Sec.~\ref{dc}. Indeed, the transverse field removes the degeneracy of the two classical symmetry-breaking ground states of the pure Ising Hamiltonian that prevents the formation of entanglement. In any case, the transverse field has to remain of moderate intensity, because a large field tends to polarize the spins and hence, again, to destroy their entanglement.

\begin{figure}[!th]
\begin{center}
\includegraphics[width=8cm]{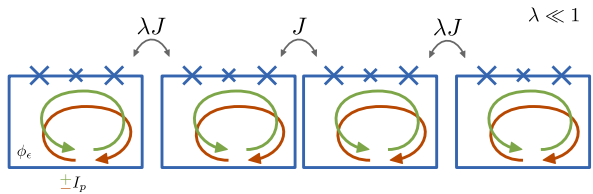}
\end{center}
\caption{An linear array of superconducting flux qubits, with a coupling pattern suitable for the simulation of LDE in quantum spin chain with open boundary conditions of the type illustrated in Fig.~\ref{figspinchain}.}
\label{figchain}
\end{figure}

In the present work we will demonstrate that the preparation of the ground state of models featuring LDE can be realized by adiabatic quantum simulation~\cite{Farhi,Grajcar05}. This technique allows for the preparation of the ground state of complex Hamiltonians by the adiabatic control of some system parameters, which allows to deform continuously the Hamiltonian from a simple configuration, whose ground state can be easily prepared, to the final target configuration. If the manipulation is slow enough, a system initialized in the ground state of the simple Hamiltonian will follow the instantaneous ground state of the evolving Hamiltonian until reaching the ground state of the final, more complex one.
A similar approach has been demonstrated recently for the simulation of LDE in a specific implementation with systems of trapped ions~\cite{ZippilliAdiab},
while in the present work we will investigate and demonstrate the adiabatic quantum simulation of LDE using instead linear arrays of superconducting flux qubits~\cite{Mooij,Orlando,Levitov,Tsomokos} interacting according to the coupling pattern illustrated in Fig.~\ref{figchain}, in order to simulate a quantum spin chain with open boundary conditions of the type reported in Fig.~\ref{figspinchain}.

We propose two protocols that are specifically engineered to make use of the simplest possible design of the superconducting circuits. The first protocol is designed to realize the adiabatic quantum simulation of an Ising model in transverse field in the regime of parameters that supports LDE. Indeed, interactions of the Ising type are naturally realized with flux qubit~\cite{Orlando}, and this is the simplest spin configuration that can be simulated with this devices, as in this protocol one needs only dc-currents in order to manipulate the qubits. The second protocol, on the other hand, makes use of a microwave fields to simulate more complex spin interactions of the $XX$ type and to adiabatically prepare the corresponding ground state. We show that for comparable preparation times of the two protocols, the latter allows for stronger end-to-end entanglement.

The paper is organized as follows. In Sec.~\ref{system} we introduce the system and discuss the general ideas for the implementation of LDE with superconducting qubits. In Sec.~\ref{dc} we present a specific protocol for the adiabatic simulation of LDE where the adiabatic manipulation is performed controlling dc-currents. Then in Sec.~\ref{mw} we discus a second protocol in which the manipulation is realized via microwave fields. The experimental feasibility is discussed and demonstrated in Sec.~\ref{discussion}. Finally, conclusions are drawn in Sec.~\ref{conclusion}.

\section{The system}\label{system}

\begin{figure}[!th]
\begin{center}
\includegraphics[width=8cm]{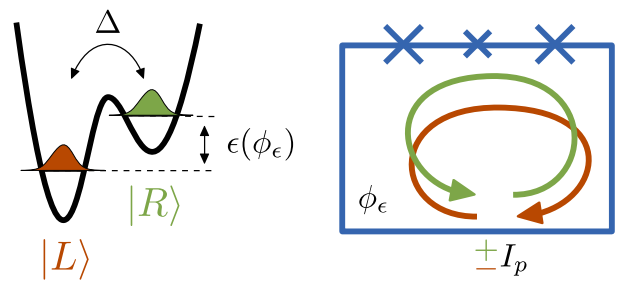}
\end{center}
\caption{The flux qubit: A superconducting loop interrupted by three Josephson Junctions is described by a double well potential where the two lowest localized states correspond to states of opposite circulating persistent currents, and constitute the base vectors of the qubit.}\label{fig1}
\end{figure}

We consider superconducting flux qubits~\cite{Mooij,Orlando}, which use states of quantized circulation (magnetic flux) in a superconducting loop interrupted by three Josephson junctions, as in Fig.~\ref{fig1}.
The dynamics of the low energy states of the system can be described by a double well potential, where the lowest localized states correspond to clockwise and anti-clockwise currents, as depicted in Fig.~\ref{fig1}.
We restrict our analysis to only these two states that constitute a base for the superconducting qubit.
We call the two states $\ke{L}$ and $\ke{R}$ respectively.
Introducing the spin operators $\sigma^z=\pt{\kb{R}{R}-\kb{L}{L}}$ and $\sigma^x=\pt{\kb{R}{L}+\kb{L}{R}}$, the corresponding Hamiltonian is analogous to that of a spin-$1/2$ particle in a magnetic field,
\begin{eqnarray}
H_{q}=-\hbar\ \epsilon\ \sigma^z-\Delta\ \sigma^x
\end{eqnarray}
where the energy difference between the localize states $\epsilon$, i.e. the magnetic energy bias,
can be controlled via the external flux $\phi_\epsilon$ threading the qubit loop, generated, for example, by a nearby dc-current line. On the other hand, $\Delta$, accounts for the tunneling amplitude between $\ke{L}$ and $\ke{R}$, and is fixed and positive.

Furthermore the flux qubit can be manipulated by microwave driving fields which modulate the energy difference between $\ke{L}$ and $\ke{R}$. Its effect can be described by an Hamiltonian of the form
\begin{eqnarray}
H_w=2\ \hbar\ \Omega\ \cos\pt{\omega\ t+\varphi}\ \sigma^z.
\end{eqnarray}

When two flux qubits are close together, they interact via their mutual inductance according to an antiferromagnetic ($J>0$) Ising Hamiltonian~\cite{Izmalkov}
\begin{eqnarray}
H_I=\hbar\ J\ \sigma^z_1\ \sigma^z_2.
\end{eqnarray}
The same antiferromagnetic as well as ferromagnetic interaction  is also obtained constructing flux qubits with a shared Josephson Junction~\cite{Grajcar,Grajcar06}.
On the other hand, more complex set up which, for example, make use of an additional dc squid as a coupler device~\cite{van der Ploeg}, allows for the interaction to be tunable.

In the present work we consider the simple situation in which $\Delta$ is fixed by construction. Nevertheless, in principle, one can imagine more complex designs which permit the control of both $\epsilon$ and $\Delta$~\cite{Fedorov}. In this case, it should be possible, in principle, to implement the same protocol discussed in Ref.~[\onlinecite{ZippilliAdiab}] for the simulation of LDE with trapped ions. However, here we aim at keeping the design as simple as possible and to analyze the performance of flux qubits as adiabatic quantum simulators with minimal control.

To be specific, we analyze the preparation of LDE with a chain of $N$ flux qubits coupled by nearest neighbor coupling, whose Hamiltonian reads
\begin{eqnarray}\label{H}
H&=&-\hbar\sum_{j=1}^N\pq{\epsilon_j\ \sigma_j^z+\Delta_j\ \sigma_j^x+2\Omega_j\ \cos\pt{\omega\ t+\phi_j}}
\nn\\&&
+\hbar\sum_{j=1}^{N-1}J_{j,j+1}\ \sigma_j^z\ \sigma_{j+1}^z\ .
\end{eqnarray}
In the following we will analyze two limiting cases
which allow to simulate two different spin models. In both cases we will show how to prepare adiabatically the corresponding ground state
and we will show that, when the end spins are weakly coupled to the bulk, such ground states exhibits LDE.

In the first place, in Sec.~\rp{dc}, we consider the case of $\Omega_j=0$,
such that the control parameters are the energy bias $\epsilon_j$.
Correspondingly, in this case we simulate an Ising spin model in transverse field,
where the role of the transverse magnetic field is played by the tunnel transition coupling, $\Delta_j$, between the superconducting states of the flux qubit.
Dc-current through a nearby current line is used to continuously tune $\epsilon_j$ from large to vanishing values. The qubits are initialized using large values of $\epsilon_j$. In this situation each flux qubit relaxes to the lowest energy eigenstate. If the noise temperature is sufficiently small then the probability to find the qubits in the excited state is negligible. In this case all the spins are polarized.
Starting from this situation the value of $\epsilon_j$ is reduced adiabatically until zero where the target Hamiltonian is realized and LDE is achieved.

On the other hand, in the second case (Sec.~\ref{mw}) the adiabatic manipulation is realized at $\epsilon_j=0$ and it is performed controlling the microwave field. In this case, an effective XX model is obtained.
Only the initialization is equal to that of the first protocol and it is realized with finite values of $\epsilon_j$ and no microwaves. Specifically, the $\epsilon_j$ are set to large values until the qubits relax to the polarized state. Then $\epsilon_j$ is set to zero and simultaneously the microwave field is switched on. As we demonstrate in Sec.~\ref{mw},
if the field drives resonantly the qubits in the regime of large $\Delta_j$, then the system dynamics approximate that of an $XX$ model in an external field whose intensity is controllable via the intensity of the driving field, which is proportional to $\Omega_j$. This intensity is therefore manipulated adiabatically: initially it is set to a sufficiently large value, such that  the initialized state is also the ground state of the effective model; then it is reduced until the ground state of the effective $XX$ model, which supports LDE, is obtained.

A final remark is in order. Although the schemes that we discuss in detail in the following sections are in principle valid for arbitrary number of spins and concern models that are exactly solvable both for finite size and in the thermodynamic limit, in this work we are mainly interested in identifying minimal protocols that can be implemented in real experiments with technological control that is currently available or in reach in the near future. In other words, we are interested in a proof-of-principle experiment with a fully controlled small-scale demonstrator, in line with the current trend in quantum simulation research. Therefore, the aim of the present work is to provide a first, clear and controlled, path towards the actual experimental simulation of LDE. It is
therefore crucial to start by considering in full detail the simplest minimal configuration of four spins for which we expect a significantly reduced experimental effort as compared to what we can expect for larger qubits chains. Stated differently, our main goal is to show that using realistic parameters the protocol can be actually realized on a minimal, fully controlled array of superconducting qubits.

Indeed, the simplest minimal situation consists of $N=4$ spins because for an odd number of spins, for example $N=3$, the ground state would be degenerate, and hence no LDE could be observed in this case. For the case $N=4$ we will study the experimental feasibility of the protocol with realistic parameters. On the other hand, the scaling of the efficiency with the number of the spins is well beyond the scope both of the present work and of the currently available quantum simulation technologies. In particular, it is worth observing that in order to optimize the adiabatic protocol for larger number of spins, it would be, most likely, necessary to consider spin models with more sophisticated coupling and entanglement patterns, such as, for instance, models supporting modular entanglement~\cite{Gualdi} or surface entanglement~\cite{ZippilliSurf}.

\section{Manipulation with DC-currents}\label{dc}

Here we study the case with no microwave driving field.
Hence we consider the simple Ising model in transverse field:
\begin{eqnarray}\label{H1}
H=-\hbar\sum_{j=1}^N\pq{\epsilon_j\ \sigma_j^z+\Delta_j\ \sigma_j^x}+\hbar\sum_{j=1}^{N-1}J_{j,j+1}\ \sigma_j^z\ \sigma_{j+1}^z
\end{eqnarray}
where $\epsilon_j$ is the tunable parameter.
We also assume, as a necessary condition for LDE, that the end spins are weakly coupled to the bulk and that the effective magnetic field of the external spins is much smaller then that of the others.
We consider antiferromagnetic couplings $J_{j,k}>0$, that are the kind of couplings that emerge naturally from the mutual inductance between two nearby flux qubits.
In particular, the simplest situation that can be realized in an  experiment consists of only four spins, such that
\begin{eqnarray} \label{parameters}
J_{2,3}&\equiv& J\ , \ \ \ \ {\rm and}\ \ \ \  J_{1,2}=J_{3,4}=\lambda\ J\ ,
\nn\\
\Delta_2=\Delta_3&\equiv&\Delta\ , \ \ \ \ {\rm and}\ \ \ \ \Delta_{1}=\Delta_4=\lambda_h\ \Delta\ ,
\nn\\
\epsilon_2=-\epsilon_3&\equiv&\epsilon\ , \ \ \ \ {\rm and}\ \ \ \ \epsilon_{4}=-\epsilon_1=\lambda_h\ \epsilon \ ,
\end{eqnarray}
with $\lambda,\lambda_h\ll 1$.
We note that, the same results that we discuss below would be valid for ferromagnetic couplings $(J_{j,k}<0)$ if the values of $\epsilon_j$ are chosen to have all the same sign.
\begin{figure}[!th]
\begin{center}
\includegraphics[width=8cm]{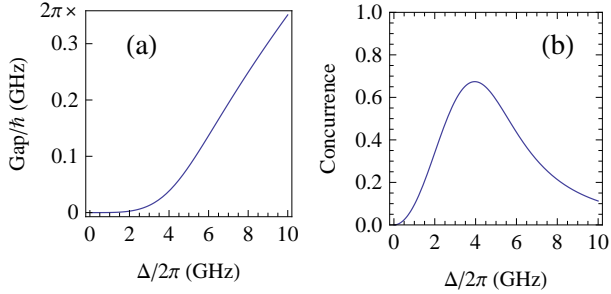}
\end{center}
\caption{(a) Gap between ground and first excited state as a function of $\Delta$, for a chain of $N=4$ flux qubits described by Eqs.~\rp{H1} and \rp{parameters}.
(b) Corresponding concurrence between first and last qubits.
The values of the parameters are $\epsilon=0$, $J=2\pi\times 5$GHz, $\lambda=0.2$, $\lambda_h=0.02$}\label{fig0}
\end{figure}

When $\epsilon=0$, this model exhibits LDE. This is shown in Fig.~\ref{fig0} where for intermediate values of the external transverse field a maximum of the concurrence (see the appendix) between first and last spins is observed. The corresponding gap between ground and first excited state, Fig~\ref{fig0} (a), is zero at vanishing $\Delta$ and increases monotonically with $\Delta$. Large gap would allow for fast adiabatic preparation, however the entanglement  is significant only for moderate values of $\Delta$ where the gap is of moderate size. This impose a trade-off between maximum velocity of the adiabatic manipulation and maximum attainable entanglement.

\begin{figure}[!th]
\begin{center}
\includegraphics[width=8cm]{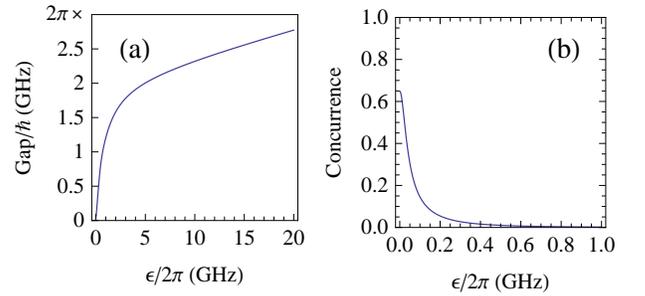}
\end{center}
\caption{(a) Gap between ground and first excited state as a function of $\epsilon$, for a chain of $N=4$ flux qubits described by Eqs.~\rp{H1} and \rp{parameters} (at $\epsilon=0$ the gap is ${\rm Gap}/\hbar\bigl |_{\epsilon=0}=2\pi\times58$MHz, which corresponds to $\sim 2.8$mK).
(b) Corresponding concurrence between first and last qubits for small values of $\epsilon$.
The values of the parameters are $\Delta=2\pi\times 4.5$GHz, $J=2\pi\times 5$GHz, $\lambda=0.2$, $\lambda_h=0.02$.
}\label{figdcAFgap}
\end{figure}

The ground state of this model is achieved by the adiabatic manipulation of the parameter $\epsilon$ from large values to zero.
The state of the system is initialized by setting  a large value of $\abs{\epsilon}$. In this situation each qubit relaxes to the lowest energy eigenstate. Specifically, if $\abs{\epsilon}$ is larger then the effective noise temperature then the probability to find a spin in the excited state is negligible. Hence, in this situation, in agreement with the staggered configuration defined in Eq.~\rp{parameters}, the spins get polarized in an antiferromagnetic state.
Then the value of $\epsilon$ is reduced according to the temporal profile
\begin{eqnarray}\label{fepsilon}
\epsilon(t)=\epsilon_0 \ \ee^{-r\ t}\ ,
\end{eqnarray}
with $r$ sufficiently small to guarantee adiabaticity.
At large $\epsilon$, in fact, the gap between ground and first excited state is large, as shown in Fig.~\ref{figdcAFgap} (a). and the adiabatic manipulation can be relatively fast. On the other hand as $\epsilon$ approaches zero the gap is significantly reduced and the velocity of the  variation have to be reduced in order to remain adiabatic. Correspondingly, as shown in Fig.~\ref{figdcAFgap} (b), maximum of the end-to-end entanglement is obtained at $\epsilon=0$.

\begin{figure}[!th]
\begin{center}
\includegraphics[width=8cm]{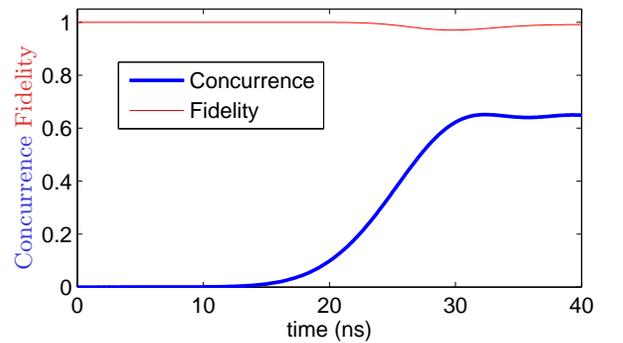}
\end{center}
\caption{End-to-end concurrence (tick, blue line) and fidelity between the evolved state and the instantaneous ground state (thin, red line) for the model described by Eqs.~\rp{H1}, \rp{parameters} and \rp{fepsilon}, when the initial state is the ground state of Eqs.~\rp{H1} and \rp{parameters} at $\epsilon=2\pi\times 20$GHz. The other parameters are
$\Delta=2\pi\times 4.5$GHz, $\epsilon_0=2\pi\times 20$GHz, $J=2\pi\times 5$GHz, $\lambda=0.2$, $\lambda_h=0.02$, $r=2\pi\times 40$MHz, N=4.}\label{fig2AF}
\end{figure}
\begin{figure}[!th]
\begin{center}
\includegraphics[width=8cm]{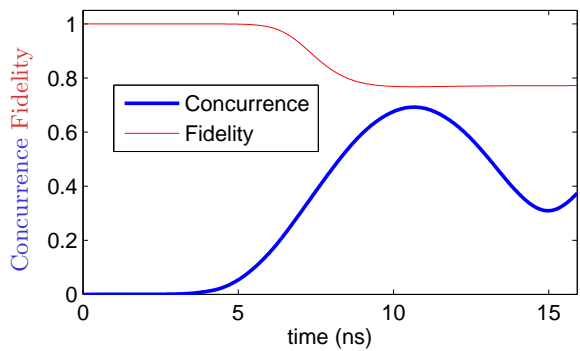}
\end{center}
\caption{As Fig.~\ref{fig2AF} with $r=2\pi\times 150$MHz.}\label{fig3AF}
\end{figure}
\begin{figure*}[!th]
\begin{center}
\includegraphics[width=14cm]{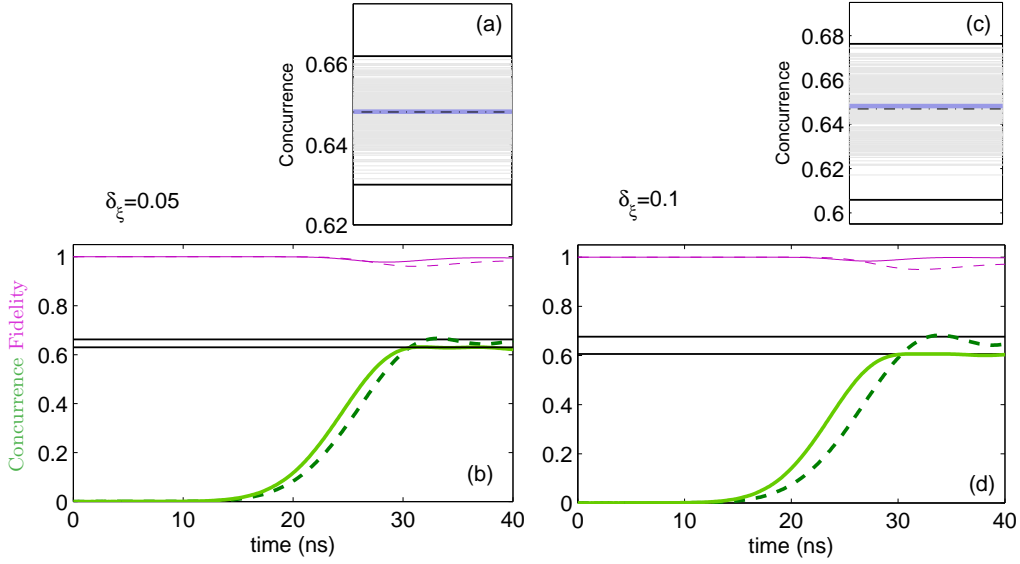}
\end{center}
\caption{Sensitivity of the protocol to random variations of the tunnel splittings $\Delta_j$. Specifically,  $\Delta_j=\Delta(1+\xi_j)$ for $j=2,3$, and $\Delta_j=\lambda_h \Delta(1+\xi_j)$ for $j=1,4$, with $\xi_j$ random variables uniformly distributed in the range $\pq{-\delta_\xi,\delta_\xi}$. Plots (a) and (b) are evaluated for $\delta_\xi=0.05$, and (c) and (d) for $\delta_\xi=0.1$. All the other parameters are as in Fig.~\ref{fig2AF}. Plots (a) and (d) report the concurrence between first and last spins, of the instantaneous ground state evaluated at the final time of the protocol, and each horizontal line correspond to a different realizations of $\xi_j$. The dashed-dotted black lines are the average concurrence over 1000 realizations, and the solid-black lines indicate the maximum and minimum realizations. The solid-thick blue lines are found for $\xi_j=0$. In (b) and (d) the horizontal solid-black lines indicate the maximum and minimum concurrence of the corresponding plots (a) and (c). The green lines and the magenta lines report, respectively,  the evolution of the end-to-end concurrence, and of the fidelity evaluated for the values of $\xi_j$ corresponding to the minimum (solid-lines) and maximum (dashed-lines) realizations of plots (a) and (c).
}\label{fig_rnd1}
\end{figure*}

The results for the adiabatic preparation are shown in Figs.~\ref{fig2AF} and \ref{fig3AF}. The thin red lines correspond to the fidelity $F=\abs{\bk{\phi[\epsilon(t)]}{\psi(t)}}^2$ between the instantaneous ground state $\ke{\phi[\epsilon(t)]}$, i.e. the ground state of the Hamiltonian for fixed values of $\epsilon(t)$ corresponding to the specific time, and the evolved sate evaluated solving the corresponding Shr\"odinger equation $\ke{\dot\psi(t)}=-\frac{\ii}{\hbar}H(t)\ke{\psi(t)}$ with the time dependent Hamiltonian given in Eq.~\rp{H1}, where the parameters are given in Eq.~\rp{parameters} and the time dependence is defined by Eq.~\rp{fepsilon}, and when the initial state is the initial ground state $\ke{\psi(0)}=\ke{\phi\pq{\epsilon(0)}}$.
When the fidelity is one, it means that perfect adiabatic following is realized. This is the case of Fig.~\ref{fig2AF}.
The corresponding concurrence between the end spins is evaluated applying the procedure discussed in the appendix to the reduced density matrix for the two spins which can be evaluated from the evolved state. We observe that in Fig.~\ref{fig2AF} the concurrence (blue thick line) saturates to a steady sizable value at large time. It means that LDE have been efficiently prepared. On the other hand, in Fig.~\ref{fig3AF} the fidelity is reduced as a result of the shorter preparation time (larger $r$). In this case the transformation is not exactly adiabatic and the system has a finite probability to be excited. The corresponding concurrence does not approach a steady value, however at finite times a strong entanglement is observed also in this case.

So far we have considered the values of $\Delta_j$ to have a well defined relative strength as defined in Eq.~\rp{parameters}. In reality, the experimental control of these parameters is nontrivial. They are fixed by construction and the actual values vary from sample to sample due to unavoidable fluctuations in the construction process. It is therefore important to analyze the sensitivity of the protocol to small deviations from the reference values set in Eq.~\rp{parameters}. The results of this analysis are reported if Fig.~\ref{fig_rnd1}, where we plot results obtained for the parameters of Fig.~\ref{fig2AF} but with $\Delta_j=\Delta\pt{1+\xi_j}$ for $j=2,3$ and $\Delta_j=\lambda_h\ \Delta\pt{1+\xi_j}$ for $j=1,4$, where $\xi_j$ are stochastic variables uniformly distributed in the range $\pq{-\delta_\xi,\delta_\xi}$. Panels (a) and (c) report the values of the end-to-end concurrence between the first and the last spin in the chain, in the instantaneous ground state corresponding to the final time of the evolution, and evaluated for $10^3$ different realizations of the variables $\xi_j$ when $\delta_\xi$ is set, respectively, to the values $0.05$ and $0.1$. The dash-dotted black lines indicate the corresponding average value, while minimum and maximum realizations are highlighted by solid black lines. The blue-solid tick line corresponds to the ideal result of Fig.~\ref{fig2AF}, namely to the case in which $\xi_j=0$ $\forall j$.

We observe that the average concurrences are very close to the corresponding ideal values obtained for $\xi_j=0$, and that the fluctuations increase with the value of $\delta_\xi$. In all cases the actual value of the concurrence remains always confined within a relatively narrow range of values. In particular, we observe that in certain cases the value of the end-to-end concurrence resulting from a random choice of the parameters can overcome that found in the initial configuration. The values of minimum and maximum realizations are also reported as black-solid horizontal lines in the plots (b) and (d). Here we use the corresponding set of parameters for the calculation of the time evolution of the concurrence (think-green lines) and of the fidelity (thin-magenta lines) under the adiabatic manipulation as done in Fig.~\ref{fig2AF}. Specifically, the solid and dashed lines correspond, respectively,  to the sets of parameters exhibiting minimum and maximum concurrence. We observe that in general the adiabatic manipulation reproduce with significant accuracy the expected values of the concurrence of the instantaneous ground-state at large time, and the fidelity is always very close to $1$. These results show that the protocol is significantly and strongly resilient to random deviations of the tunnel splitting $\Delta_j$.

\section{Manipulation with microwave fields}\label{mw}

In this section we show how to simulate $XX$ sin-1/2 models with flux qubits. This is realized by means of a microwave field driving the qubits in the regime of large $\Delta$.
In this case the Hamiltonian reads
\begin{eqnarray}\label{Ht}
H(t)&=&-\hbar\sum_{j=1}^N\pq{\epsilon_j\ \sigma_j^z+2\Omega \cos\pt{\omega\ t+\varphi_j}\ \sigma_j^z+\Delta_j\ \sigma_j^x}
\nn\\&&
+\hbar\sum_{j=1}^{N-1}J_{j,j+1}\ \sigma_j^z\ \sigma_{j+1}^z
\end{eqnarray}
where the value of $\Omega$ is assumed to be the same for all the spins.
At the anticrossing point, that is when  $\epsilon_j=0$, it is useful to study the system in
interaction picture defined by the unitary transformation
\begin{eqnarray}\label{U0}
U_0=\ee^{-\ii H_0 t}
\end{eqnarray}
where $H_0=-\sum_{j=1}^N\frac{\omega}{2}\ \sigma_j^x,$
and such that the transformed state $\ke{\widetilde\psi}=U_0\da\ke{\psi}$, satisfies $\frac{\partial}{\partial t}\ke{\widetilde\psi}=-\ii \widetilde H(t)\ke{\widetilde\psi}$
with
\begin{eqnarray}\label{tildeH}
\widetilde H(t)&=&U_0\da\ H(t)\ U_0-H_0
\nn\\
&=& -\hbar\sum_{j=1}^N\pq{\Omega\ \pt{\cos{\varphi_j}\ \sigma_j^z+\sin{\varphi_j}\ \sigma_j^y}+\pt{\Delta_j-\frac{\omega}{2}} \sigma_j^x}
\nn\\&&
+\hbar\sum_{j=1}^{N-1}\frac{J_{j,j+1}}{2} \pt{\sigma_j^y\sigma_{j+1}^y+\sigma_j^z\sigma_{j+1}^z}
\nn\\&&-
\hbar\sum_{j=1}^N\Omega\ \pq{\sin\pt{2\omega\ t+\varphi_j} \sigma_j^y+\cos\pt{2\omega\ t+\varphi_j} \sigma_j^z }
\nn\\&&
+\hbar\sum_{j=1}^{N-1}\frac{J_{j,j+1}}{2}
\lpq{\cos\pt{2\omega\ t} \pt{\sigma_j^z\sigma_{j+1}^z-\sigma_j^y\sigma_{j+1}^y}
}\nn\\&&\rpq{
+\sin\pt{2\omega\ t} \pt{\sigma_j^y\sigma_{j+1}^z+\sigma_j^z\sigma_{j+1}^y}
}\ .
\end{eqnarray}
Let us now assume that the tunnel splittings are the same for all the qubits, $\Delta_j=\Delta$ $\forall j$.
If we set the field to be resonant with the energy gap between the single spin energy enigenstates, $\omega=2\Delta$, and  we neglect the fast oscillating terms  under the assumption $4\Delta\gg\Omega,{J_{j,k}}/{2}$, Eq.~\rp{tildeH} can be approximated by the effective time independent Hamiltonian
\begin{eqnarray}\label{Heff}
\widetilde H_{eff}&=& -\hbar\sum_{j=1}^N\Omega\ \pt{\cos{\varphi_j}\ \sigma_j^z+\sin{\varphi_j}\ \sigma_j^y}
\nn\\&&
+\hbar\sum_{j=1}^{N-1}\frac{J_{j,j+1}}{2} \pt{\sigma_j^y\sigma_{j+1}^y+\sigma_j^z\sigma_{j+1}^z},
\end{eqnarray}
which describes an $XX$ spin chain in an external magnetic field whose intensity is proportional to $\Omega$.

As in the previous section, we consider only four spins, with the couplings of the end spins scaled by the small factor $\lambda$. Moreover, also in this case, we assume antiferromagnetic interactions, and correspondingly the phases $\varphi_j$ should be opposite for neighboring spins. Hence we assume
\begin{eqnarray}\label{parameters2}
J_{2,3}&\equiv& J\ , \ \ \ \ {\rm and}\ \ \ \  J_{1,2}=J_{3,4}=\lambda\ J\ ,
\nn\\
\varphi_2=\varphi_4&=&0\ , \ \ \ \ {\rm and}\ \ \ \ \varphi_1=\varphi_3=\pi\ ,
\end{eqnarray}
with $\lambda\ll 1$.
The following results apply also to the ferromagnetic case when $\varphi_j=0$ $\forall j$.

\begin{figure}[!th]
\begin{center}
\includegraphics[width=8cm]{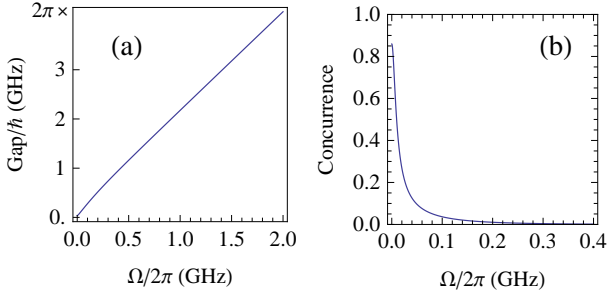}
\end{center}
\caption{(a) Gap between ground and first excited state as a function of $\Omega$, for a chain of $N=4$ flux qubits described by the effective model in Eqs.~\rp{Heff} and \rp{parameters2} (at $\Omega=0$ the gap is ${\rm Gap}/\hbar\bigl |_{\Omega=0}=2\pi\times38$MHz, which corresponds to $\sim 1.8$mK).
(b) Corresponding concurrence between first and last qubits.
The values of the other parameters are $J=2\pi\times 1$GHz, $\lambda=0.2$.}\label{figMWgap}
\end{figure}
\begin{figure}[!th]
\begin{center}
\includegraphics[width=8cm]{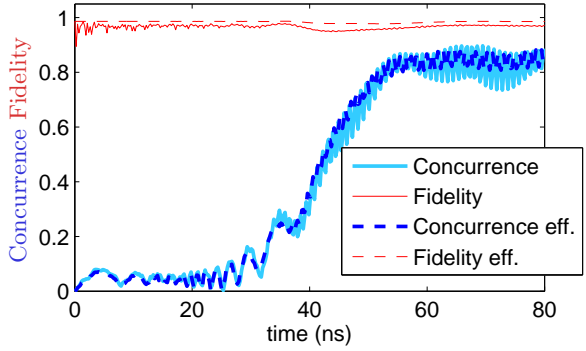}
\end{center}
\caption{End-to-end concurrence (tick, blue lines) and fidelity between the evolved state and the instantaneous ground state (thin, red lines) for the full model described by Eq.~\rp{Ht} with $\epsilon_j=0$ (solid lines), and for the effective model described by Eq.~\rp{Heff} (dashed lines), when the initial state is
the ground state of Eqs.~\rp{Ht} at $\Omega=0$ and $\epsilon_{j=2,4}=-\epsilon_{j=1,3}=2\pi\times 100$GHz.
In all cases we consider a chain of $N=4$ spins with the parameters defined in Eqs.~\rp{parameters2} and \rp{fOmega}.
The values of the other parameters are $\Delta_j=\omega/2=2\pi\times 10$GHz $\forall j$, $\Omega_0=2\pi\times 2$GHz, $J=2\pi\times 1$GHz, $\lambda=0.2$, $r=2\pi\times 0.02$GHz.}\label{fig4}
\end{figure}
\begin{figure}[!th]
\begin{center}
\includegraphics[width=8cm]{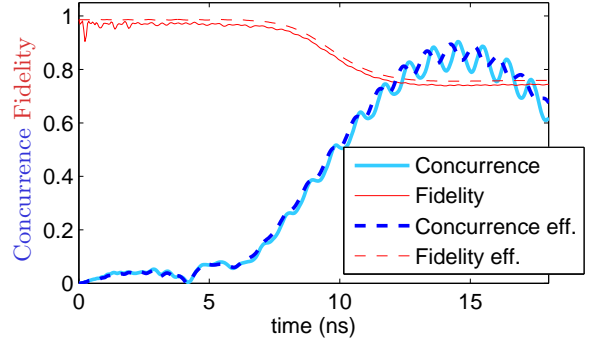}
\end{center}
\caption{As Fig.~\ref{fig4} with $r=2\pi\times 0.1$GHz.}\label{fig5}
\end{figure}
\begin{figure*}[!th]
\begin{center}
\includegraphics[width=14cm]{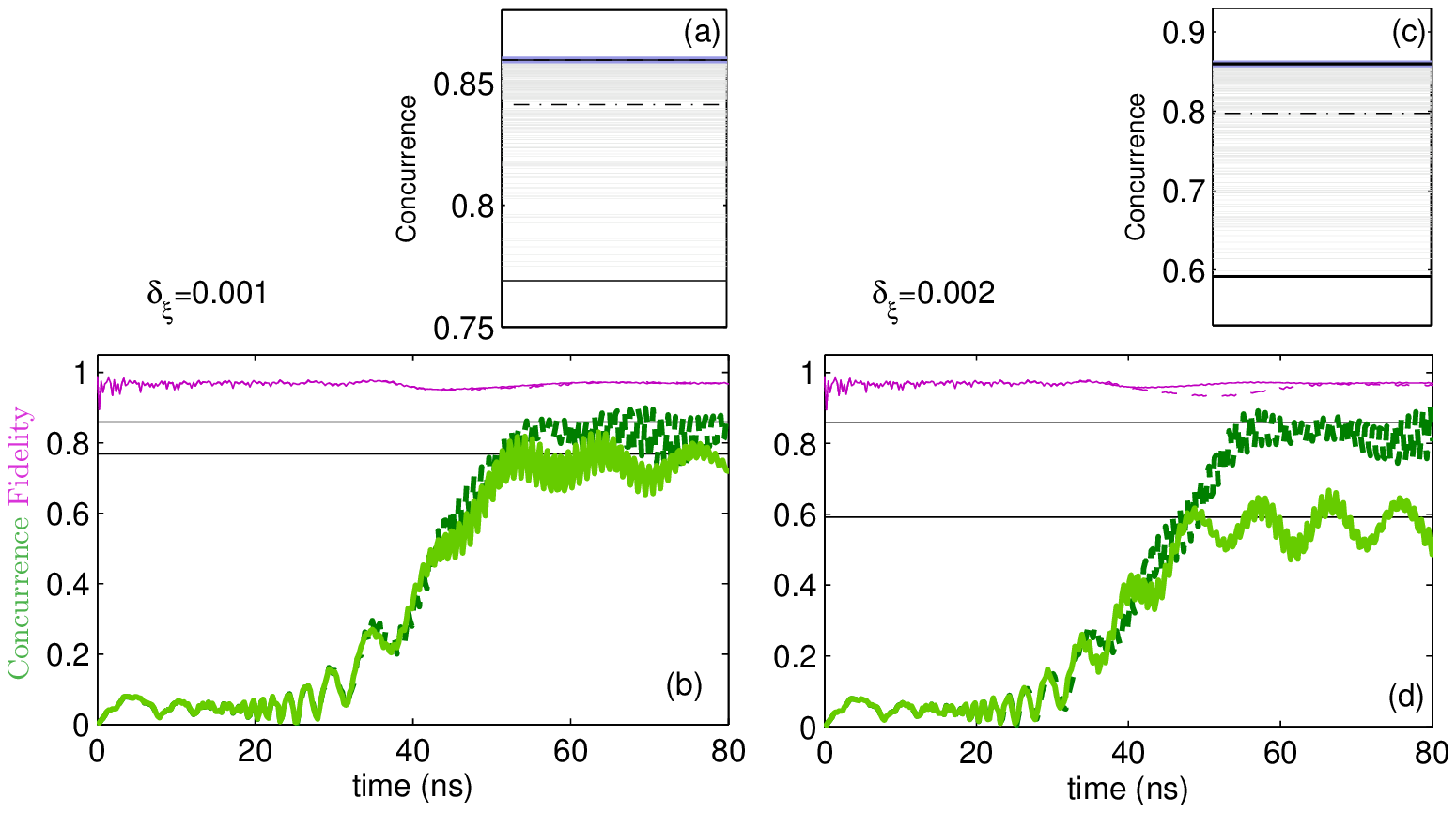}
\end{center}
\caption{
Sensitivity of the protocol to random variations of the tunnel splittings $\Delta_j$. Specifically,  $\Delta_j=\Delta(1+\xi_j)$, with $\xi_j$ random variables uniformly distributed in the range $\pq{-\delta_\xi,\delta_\xi}$. Plots (a) and (b) are evaluated for $\delta_\xi=0.001$, and (c) and (d) for $\delta_\xi=0.002$. All the other parameters are as in Fig.~\ref{fig4}. Plots (a) and (d) report the concurrence between first and last spins, of the instantaneous ground state of $\tilde H_{eff}+H_\xi$ (with $\tilde H_{eff}$ defined in Eq.~\rp{Heff} and $H_\xi$ in Eq.~\rp{Hxi}) evaluated at the final time of the protocol, and each horizontal line correspond to a different realizations of $\xi_j$. The dashed-dotted black lines are the average concurrence over 1000 realizations, and the solid-black lines indicate the maximum and minimum realizations. The solid-thick blue lines are found for $\xi_j=0$. In (b) and (d) the horizontal solid-black lines indicate the maximum and minimum concurrence of the corresponding plots (a) and (c). The green lines and the magenta lines report, respectively,  the evolution of the end-to-end concurrence, and of the fidelity evaluated, with the total Hamiltonian in Eq.~\rp{Ht}, for the values of $\xi_j$ corresponding to the minimum (solid-lines) and maximum (dashed-lines) realizations of plots (a) and (c).
}\label{fig_rnd2}
\end{figure*}

The adiabatic preparation goes as follows. First the system has to be initialized. This step is similar to the one already discussed in the protocol of Sec.~\ref{dc}. We consider no microwave field $\Omega=0$ and we set the single spin	enrgies $\epsilon_j$ to large values (much larger then $\Delta$) in a staggered configuration $\epsilon_1=\epsilon_3=-\epsilon_2=-\epsilon_4$.
The system then relax to the ground state, that is the anti-ferromagnetic state, if the effective noise temperature is smaller then $\abs{\epsilon_j}$.
Once the system is initialized, the parameters $\epsilon_j$ are set to zero and the microwave field is switched on with an amplitude, $\Omega$, larger than $J_{j,k}/2$ (note however that $\Omega$ have to be smaller than $\Delta$ in order for the effective model to be valid). In this way, if the phases of the driving fields are set as in Eq.~\rp{parameters2}, then the initialized state is also ground state of the effective Hamiltonian.   In particular while in the original frame the spins will start to oscillate under the effect of the driving field, in the transformed reference frame the antiferromagnetic state will remain stationary.

At this point, the amplitude of the driving field $\Omega$ is manipulated adiabatically from large values to zero according to the temporal function
\begin{eqnarray}\label{fOmega}
\Omega(t)=\Omega_0 \ \ee^{-r\ t}\ .
\end{eqnarray}
Thereby, using a sufficiently small $r$, the state of the system follows the instantaneous ground state until it approaches the ground state of the $XX$ model which exhibits LDE.
We note that the adiabatic following takes place effectively in the transformed reference frame, nevertheless,
since the transformation $U_0$ is local, the entanglement properties in the two representation are the same.
Meaning that also in the original representation the end spins get entangled.

As shown in Fig.~\ref{figMWgap} (a) also in this case the gap is large for large values of the effective magnetic field, in this case $\Omega$, and decreases as $\Omega$ decreases. Hence, similarly to the first protocol, see Eq.~\rp{fepsilon},  the adiabatic transformation can be fast at the beginning and have to slow down when the gap reduces.
In particular we have chosen the system parameters such that the corresponding gap is of the same order of magnitude of the one presented in Fig.~\ref{figdcAFgap} for the first  protocol. In this way we expect that the preparation time for the two protocols is similar. On the other hand, we observe that the achievable end-to-end concurrence is significantly larger in this case.

Numerical results for this protocol are presented in Figs.~\ref{fig4} and \ref{fig5}.
In this case we compare two cases: the results obtained integrating the Schr\"odinger equation with the full Hamiltonian in Eq.~\rp{Ht} (solid lines) and that obtained with the effective Hamiltonian in Eq.~\rp{Heff} (dashed lines).
The red thin lines depict the fidelity between the instantaneous ground state, $\ke{\widetilde\phi\pq{\Omega(t)}}$, of the effective Hamiltonian~\rp{Heff}, and the evolved states under the two Hamiltonians corresponding to the solid and dashed lines respectively. We note that  the evolved states $\ke{\psi(t)}$ under the full Hamiltonian~\rp{Ht}, can be compared to the instantaneous ground state, $\ke{\widetilde\phi\pq{\Omega(t)}}$, only after the latter is transformed back to the original representation by the application of the unitary transformation $U_0$ defined in Eq.~\rp{U0}. The two states are in fact defined in two different representations, therefore the corresponding fidelity (thin red solid lines) is found as $F=\abs{\br{\widetilde\phi\pq{\Omega(t)}}U_0\da\ke{\psi(t)}}^2$.
On the other hand the fidelity for the evolved state $\ke{\widetilde\psi_{eff}(t)}$ under the effective Hamiltonian~\rp{Heff} (thin red dashed lines) is found as $F=\abs{\bk{\widetilde\phi\pq{\Omega(t)}}{\widetilde\psi_{eff}(t)}}^2$.
The numerical results show that the fidelity for both the full and the effective model are always very close meaning that the system of flux qubits well simulates the effective $XX$ spin model. In particular in Fig.~\ref{fig4} the fidelity is always very close to one indicating that almost perfect adiabatic following is realized under this conditions.
On the other hand when the preparation time is reduced (the rate $r$ appearing in Eq.~\rp{fOmega} is increased) as in Fig.~\ref{fig5} then the system can be excited and the fidelity is reduced.

The corresponding end-to-end concurrence is obtained applying the definition given in the appendix to the reduced density matrix for the end spins which, in turn, is evaluated form the evolved states $\ke{\psi(t)}$ (thick solid light-blue line) and $\ke{\widetilde\psi_{eff}(t)}$ (thick dashed dark-blue line). The results for the two models are always very similar confirming that the system simulate accurately the $XX$ chain. In particular, in full similarity with the protocol of Sec.~\ref{dc}, when the adiabatic condition is fulfilled as in Fig.~\ref{fig4} the concurrence approaches a large steady value; and,  when the manipulation is too fast, the entanglement at large time oscillate, however reaching large values at finite times.

Finally, also in this case we analyze the stability of the protocol to random deviations of the tunnel splitting $\Delta_j$ from the ideal condition of equal values assumed so far. Moving along lines similar to the analysis presented in the previous section, we now consider the parameters of Fig.~\ref{fig4}, and we assume $\Delta_j=\Delta(1+\xi_j)$ with $\xi_j$ random variable uniformly distributed in the range $\pq{-\delta_\xi,\delta_\xi}$.
In this case, in Fig.~\ref{fig_rnd2} (a) and (c), we report the final end-to-end concurrence evaluated from the instantaneous ground state of the effective Hamiltonian in Eq.~\rp{Heff} with an added term of the form
\begin{eqnarray}\label{Hxi}
H_\xi=\Delta \sum_j \xi_j\ \sigma_j^x.
\end{eqnarray}
On the other hand, the time evolutions for the concurrence (green tick lines) and the fidelity (magenta thin lines) in Fig.~\ref{fig_rnd2} (b) and (d) are evaluated from the full Hamiltonian in Eq.~\rp{Ht}.

As in the previous discussions, also in this case we consider the maximum and minimum realizations, as shown in panels (a) and (c), and we report them, as horizontal black lines, in panels (b) and (d). The corresponding parameters are then used to evaluate the time evolutions in panels (b) and (d).
We observe once more that at large time the end-to-end concurrence approaches the corresponding expected results (horizontal black lines), and the fidelity is always close to one, meaning that the adiabatic simulation scheme continues to work also in the presence of fluctuations in the tunnel splitting.
However, at variance with the first protocol, we observe from Fig.~\ref{fig_rnd2}, panels (a) and (c), that deviations from the initial configuration tend always to reduce the end-to-end concurrence. Moreover, this reduction can be significant already at quite small values of $\delta_\xi$. This is due to the fact that the ideal situation is found when $\xi_j\sim0$, i.e. when $\xi_j\Delta$ is much smaller then the other parameters, in particular it means that the optimal situation is found when $\xi_j\Delta\ll J$. However, we have seen that this schemes works under the condition $J\ll \Delta$. These facts imply that the second protocol operates properly only when $\xi_j\ll J/\Delta\ll 1$, hence requiring an higher degree of experimental control over the system parameters as compared to the first protocol.

\section{Experimental realizability}\label{discussion}

The presented results demonstrate that superconducting flux qubits can be used to simulate spin models which exhibit LDE.
In our analysis we have completely disregarded the effect of dephasing and relaxation on the system dynamics.
This is justified if the preparation time is much smaller then the time scales for dephasing and relaxation of flux qubits which, otherwise, would inevitably destroy the quantum coherences that we aim to observe.

The largest dephasing and relaxation times reported to date are of the order of few $\mu$s~\cite{Bylander,Stern}.
Our results, on the other hand, demonstrate that LDE can be prepared in a time of the order of $10$ ns, which is, hence, much shorter than the typical dephasing times in standard flux qubits experiments.
The preparation time is mainly constrained by the value of the spin-spin coupling constant $J$, where larger $J$ allows for faster preparation. We have obtained our results with spin-spin coupling constants as large as $J=2\pi\times 5$GHz which seems a very reasonable value already achieved in experiments~\cite{Grajcar,Grajcar06}. Larger $J$, that can be realized by Josephson junction coupling~\cite{Grajcar}, would allow to further reduce the preparation time. Moreover, speeding up the preparation can also be achieved by optimizing the adiabatic manipulation. One can, for example, employ techniques of optimal coherent control~\cite{Werschnik} to find the optimal time profile of the control parameters. Or one can, also, implement techniques based on shortcuts to adiabaticity~\cite{Torrontegui,Bason,Opatrny} to optimize the effect of additional control parameters.
These observations seems to suggest that these protocols are not relevantly affected by realistic noise and hence are very promising for an actual preparation of LDE.
We further remark that the effect of dephasing can be in principle reduced by employing decoherence control techniques such as dynamical decoupling in a similar fashion of that discussed in Ref.~[\onlinecite{Bylander}].

We finally comment about the readout of the resulting entanglement. The preparation and measurement of up to three-qubit entanglement has already been achieved in superconducting circuits~\cite{DiCarlo10}. Similar detection techniques should be applied in our case to analyze LDE.
More in general, in order to prove that the state of two spins is entangled one can show, for example, that it violates the Bell's inequality~\cite{Ansmann}. One can also provide a full characterization of the two spin state by quantum state tomography~\cite{Liu,Steffen,Filipp}. In any case one should identify strategies to measure various correlation function of spin operators along different directions. Specifically, LDE can be probed by the measurement of only first and last qubit.

The measurement of the state of flux qubits can be realized, for example, using a dc-squid~\cite{Fedorov,Bylander}, by dispersive readout using high-quality superconducting resonators coupled to the qubit~\cite{Greenberg02, Grajcar04, Born04, Wallraff04,Izmalkov}, or by their combination \cite{Lupascu05, Reuther11}.
In particular, the dispersive readout measures the observable $\sigma_x$ while the dc-squid measures $\sigma_z$. Both techniques require single spin rotations to reconstruct the spins state by quantum sate tomography~\cite{Steffen}.
On the other hand their combination allows to select the measured observable by the control of external parameter (bias current or magnetic flux)\cite{Reuther11}. In this way one can avoid single spin rotations which prolong the measurement time and would require very fast pulses.

\begin{figure}[!th]
\begin{center}
\includegraphics[width=8cm]{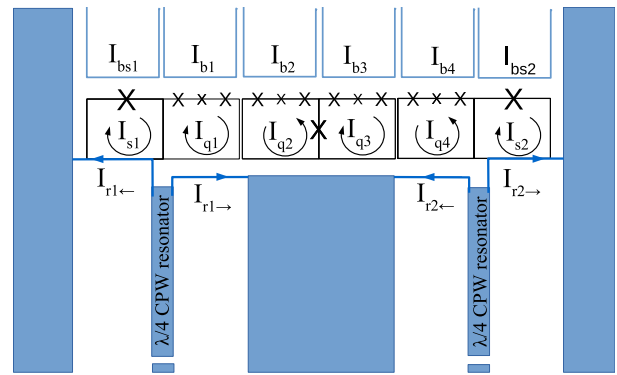}
\end{center}
\caption{Scheme of a chain of $4$ flux qubits with two rf-SQUIDs which perform the dispersive readout of end qubits. The individual flux qubits are biased by dc currents $I_{bi}$, with $i={1,2,3,4}$ . The rf currents
flowing in the middle wires of two $\lambda/4$ coplanar waveguide resonators are split into the currents $I_{ri\leftarrow}$ and $I_{ri\rightarrow}$ (with $i=1,2$) flowing to the left and to the right ground plane, respectively. The currents $I_{r1\leftarrow}$ and $I_{r2\rightarrow}$ flow through the superconducting  wires, with inductance $L_w$, shared with the rf-SQUIDs, which are coupled to the end qubits 1 and 4, respectively.
The rf-SQUIDs mediates an interaction between the qubits and the resonantor whose resonance frequency is therefore sensitive to the state of the qubits. The currents $I_{bs1},I_{bs2}$ biasing the rf-SQUIDs  are used to control the kind of coupling between qubits and  rf-SQUIDs, and in turn they allow to control the sensitivity of the resonance frequency of the resonators to the observables $\sigma_z$ and $\sigma_x$.
}\label{fig:ResSquid}
\end{figure}

The detection devices, i.e. dc-SQUID and/or superconducting resonators, should be included in the design of the superconducting circuit and should be constructed in order to allow for the detection of the two end spins.
A possible detection scheme which would allow to perform quantum tomography of the two end qubits state is reported in Fig.~\ref{fig:ResSquid}.
In this case, each qubits is probed by an rf-SQUID which, in turn, is coupled to a superconducting resonator.
The dc-magnetic fluxes produced by two independent wires bias the rf-SQUIDs to the working point which determine the sensitivity of the resonator to the observable $\sigma_z$ and $\sigma_x$\cite{Reuther11}.
%

\begin{figure}[!th]
\begin{center}
\includegraphics[width=8cm]{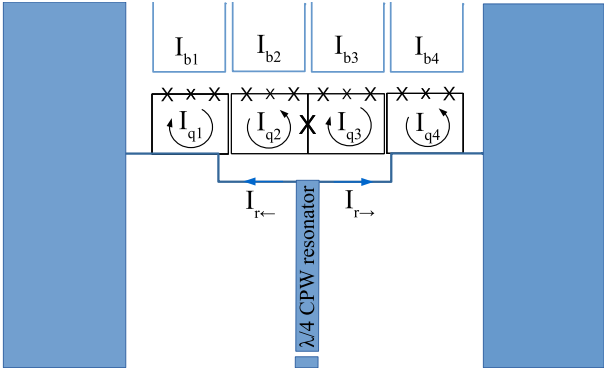}
\end{center}
\caption{Scheme of a chain of $4$ flux qubits with joint dispersive readout of the end qubits with one resonator. The individual flux qubits are biased by dc currents $I_{bi}$, with $i=1,2,3,4$. The rf current
flowing in the middle wire of a $\lambda/4$ coplanar waveguide resonator is split into the currents $I_{r\leftarrow}$ and $I_{r\rightarrow}$ flowing to the left and to the right ground plane, respectively. These currents
flow through the superconducting  wires, with inductance $L_w$, shared with end qubits 1 and 4, respectively.  It provides the coupling constant
$\kappa\approx L_w/\sqrt{L_qL_r}$ between the resonator with inductance $L_r$ and end qubits wih inductance $L_q$.}\label{fig:joint}
\end{figure}

An alternative possibility is to use a joint qubits readout scheme for the direct detection of the state of the two qubits~\cite{Smirnov03,Izmalkov,Chow10}.
This kind of measurement  can be realized with a device similar to the one shown in Fig.~\ref{fig:joint}.
%
The two weakly coupled qubits at the ends of the chain are coupled to a single superconducting resonator. The resonance of the resonator is detuned from the qubits transition frequency in order to achieve the dispersive regime.

In this regime, the detuning of the resonance frequency of the combined resonator-qubit system depends on the state of the coupled qubits according to the following relation~\cite{Smirnov03,Izmalkov}:
\begin{equation}
\frac{\Delta\omega_r}{\omega_r}=\kappa^2R_{ge}\frac{L_qI_q^2}{\Delta E} \; ,
\end{equation}
where $\Delta E$ is the energy gap between the ground state and the first excited energy level, $\kappa$ is the coupling constant between qubits and resonator, $L_q$ is the inductance of the qubit, $I_q$ is the persistent current of the qubit, and $R_{ge}$ is the real matrix element defined as:
\begin{eqnarray}
R_{ge}&=& \av{g|\sigma^z_1|e}\av{e|\sigma^z_1|g}+
          \av{g|\sigma^z_4|e}\av{e|\sigma^z_4|g}\nonumber \\
      &-&\av{g|\sigma^z_1|e}\av{e|\sigma^z_4|g}-
        \av{g|\sigma^z_4|e}\av{e|\sigma^z_1|g} \; .
        \label{eq:R12}
\end{eqnarray}
If the qubits are not entangled, the shift of the resonance frequency is only determined by the first two terms in Eq.~\ref{eq:R12}, which are clearly positive and nonvanishing. On the other hand, an entangled state gives additional contributions to the shift expressed by the last two terms. These describe a coherent flipping of both qubits, which is possible only for non-factorizable eigenstates. Thereby, one can probe the entanglement between the qubits by comparing the transmission of microwaves with angular frequency $\omega_r$ through the resonator before and after the adiabatic preparation of LDE.

In this kind of measurements, the measurement time is determined by the quality factor $Q$ of the resonator and can be estimated along the lines put forward in Ref.~\cite{Tornberg07}:
%
\begin{equation}
t_{meas}\sim\max{\{\frac{k_BT_N}{L_qI_q^2}\frac{1}{\kappa^2Q^2\omega_r},\frac{Q}{\omega_r}\}} \; ,
\label{eq:tmeas}
\end{equation}
where $k_B$ is the Boltzmann constant and $T_N$ is the noise temperature of the amplifier.
For typical parameters of superconducting qubits $L_q\approx25$~pH, $I_q\approx 0.25$ $\mu$A, $\kappa\approx0.01$, microwave cryogenic amplifiers $T_N\approx 5$~K and
superconducting resonators $\omega_r=2\pi\times 7.5$~GHz   the measurement time is minimized for $Q\approx75$ (see Eq.~\ref{eq:tmeas}) providing the measurement time $t_{meas}\approx 1.5$~ns.
This value is much shorter then the preparation time. It is, therefore, promising for an actual detection of LDE also in the non-optimal conditions, in which the final state of the system is not stationary, as that described in Figs.~\ref{fig3AF} and \ref{fig5}.

\section{Conclusions}\label{conclusion}

We have demonstrated that the ground-state long-distance entanglement (LDE) featured in certain one-dimensional quantum spin models with open boundaries can be efficiently realized with linear arrays of superconducting flux qubits. We have analyzed two protocols corresponding, respectively, to two different quantum spin chains featuring LDE.
The first one requires only the use of dc-currents to engineer the system dynamics and realizes the quantum simulation of an open Ising chain in transverse field. The second protocol exploits microwave fields to simulate open $XX$ chains with competing interactions along two different components of the spin. Indeed, under comparable preparation times, the latter results in a larger end-to-end concurrence. For both protocols we have shown that, assuming realistic parameters, the preparation time is relatively short. Specifically, it is much shorter then the typical dephasing time in this type of systems. This suggests that the protocols are to a large extent insensitive to realistic sources of noise. A very challenging, but fascinating, future research direction would be the generalizations of these protocols to the simulation of symmetry-protected topological order in more complex one-dimensional systems and the quantum simulation of surface entanglement, the quite rich two-dimensional analogue of one-dimensional long-distance entanglement.

\section*{Acknowledgments}
All authors acknowledge funding by the European Union Seventh Framework Programme (FP7/2007-2013) under grant agreement No. 270843 (FP7 STREP Project iQIT). M.G. acknowledges partial support of Slovak Research and Development Agency Contract No. DO7RP-0032-11, APVV-0515-10 and APVV-0808-12.
E.I. acknowledges a partial support by the Russian Ministry of Science and education, Contract No. 8.337.2014/K.
\appendix
\section{Concurrence}\label{concurrence}

The concurrence is a measure of the entanglement between two qubits, directly related to the entanglement of formation. Concurrence is maximum (reaches unity) in the pure, maximally entangled Bell states.

Given a density matrix  for two spins $\rho$,
the corresponding concurrence $\CC(\rho)$ is computed applying the definition
\begin{eqnarray}
\CC(\rho)={\rm max}\pg{0,\lambda_1-\lambda_2-\lambda_3-\lambda_4}
\end{eqnarray}
where $\lambda_1,...\lambda_4$ are the eigenvalues in decreasing order of the matrix $\Lambda=\sqrt{\rho\ \tilde\rho}$
with
\begin{eqnarray}
\tilde\rho=\sigma^y\otimes \sigma^y\ \rho^*\ \sigma^y\otimes \sigma^y\
\end{eqnarray}
and where $\sigma^y$ is the Pauli spin matrix and the symbol $^*$ indicates the complex conjugations of the elements of $\rho$.

\end{document}